# Mortality decrease and mathematical limit of longevity

Byung Mook Weon

LG.Philips Displays, 184, Gongdan1-dong, Gumi-city, GyungBuk, 730-702, South Korea

**Abstract**

We wish to verify that the mortality deceleration (or decrease) is a consequence of the bending of the shape parameter at old ages. This investigation is based upon the Weon model (the Weibull model with an age-dependent shape parameter) for human survival and mortality curves. According to the Weon model, we are well able to describe the mortality decrease after the mortality plateau, including the mortality deceleration. Furthermore, we are able to simply define the mathematical limit of longevity by the mortality decrease. From the demographic analysis of the historical trends in Switzerland (1876-2001) and Sweden (1861-2001), and the most recent trends in the other eleven developed countries (1996-2001), we confirm that the bending of the shape parameter after characteristic life is correlated with the mortality deceleration (or decrease). As a consequence, this bending of the shape parameters and the mortality deceleration is associated with the mathematical limit on longevity. These results suggest that the mathematical limit of longevity can be induced by the mortality deceleration (or decrease) in nature. These findings will give us a breakthrough for studying the mortality dynamics at the highest ages.

# 1. Introduction

Fundamental studies of the aging process have lately attracted the interest of researchers in a variety of disciplines, linking ideas and theories from such diverse fields as biochemistry to mathematics (Weitz and Fraser 2001). The way to characterize aging is to plot the increase in mortality rate with chronological age. The mortality rate is the probability that an individual who is alive at a particular age will die during the following age interval. The mortality rate can also be represented as the fraction of the population surviving to a particular age (or the survival rate).

The fundamental law of population dynamics is the Gompertz law (Gompertz 1825), in which the human mortality rate increases roughly exponentially with increasing age at senescence. The Gompertz model is most commonly employed to compare mortality rates between different populations (Penna and Stauffer 1996). However, no mathematical model so far, including the Gompertz model, has been suggested that can perfectly approximate the development of the mortality rate over the total life span (Kowald 1999). Particularly in modern research findings, it seems to be obvious that the mortality rate does not increase according to the Gompertz model at the highest ages (Vaupel 1997, Robine and Vaupel 2002), and this deviation from the Gompertz model is a great puzzle to demographers, biologists and gerontologists. There are two standard hypotheses that have been put forward to explain this phenomenon: the individual-risk hypothesis and the heterogeneity hypothesis (Higgins 2003). This puzzle needs to be resolved.

There is strong evidence from many developed countries that the rate of increase in mortality decelerates at high ages. However, many of the traditional mathematical models (for instance, the Gompertz, Weibull, Heligman & Pollard, Kannisto, Quadratic and Logistic models) for the mortality rate provide poor fits to empirical population data at the highest ages (Thatcher, Kannisto, and Vaupel 1998, Yi and Vaupel 2003). We have recently found a useful model derived from the Weibull model with an age-dependent shape parameter to describe the human survival and mortality curves (Weon 2004a, 2004b). The model suggests that that the mortality deceleration (or decrease) at old ages is a consequence of the bending of the shape parameter (Weon 2004a). We are able to apply the model to describe the mortality deceleration (or decrease) at older ages. In this investigation, we wish to demonstrate the mortality deceleration (or decrease) by the bending of the shape parameter through the demographic analysis of the historical trends in Switzerland (1876-2001) and Sweden (1861-2001), and the most recent trends in the other eleven developed countries (1996-2001). In fact, the mortality rate decelerates at higher ages, reaching perhaps a maximum or ceiling around age 110 (Vaupel et al. 1998, Helfand and Inouye 2002). We will show that the new model enables us to describe the mortality dynamics (deceleration and/or decrease) at the highest ages.

**2. Weon model: Weibull model with age-dependent shape parameter**

We have put forward a general expression for human survival and mortality rate in our previous papers (Weon 2004a, 2004b). It is recently discovered that human survival and mortality curves are well described by the following new mathematical model, derived from the Weibull survival function and it is simply described by two parameters, the age-dependent shape parameter and characteristic life,

$$S = \exp(-(t/\alpha)^{\beta(t)}) \qquad (1)$$

where $S$ denotes the survival probability of surviving to age $t$, $\alpha$ denotes characteristic life and $\beta(t)$ denotes the shape parameter as a function of age. The original idea was obtained as follows: typical human survival curves show i) a rapid decrease in survival in the first few years of life and ii) a relatively steady decrease and then an abrupt decrease near death thereafter (see Fig. 1 in Weon 2004a, Azbel 2002). Interestingly, the former behaviour resembles the Weibull survival function with $\beta<1$ and the latter behaviour seems to follow the case of $\beta \gg 1$. With this in mind, it could be assumed that shape parameter is a function of age. The new model is completely different from the traditional Weibull model in terms of 'age dependence of the shape parameter'. It is especially noted that the shape parameter '$\beta(t)=\ln(-\ln S)/\ln(t/\alpha)$' can indicate a 'rectangularity' of the survival curve. The reason for this is that as the value of the shape parameter becomes a high value, the shape of the survival curve approaches a further rectangular shape. From now on, we will call the new model the 'Weon model'.

The Weibull model for technical devices has a constant shape parameter, whereas the Weon model for humans has an age-dependent shape parameter.

$$S = \exp(-(t/\alpha)^\beta) \quad \text{--- technical devices : Weibull model}$$

$$S = \exp(-(t/\alpha)^{\beta(t)}) \quad \text{--- humans : Weon model}$$

The shape parameter for humans, '$\beta(t)=\ln(-\ln S)/\ln(t/\alpha)$', is a function of age ($t$).

Therefore, the mortality function is described by the mathematical relationship with the survival function ($\mu = -d \ln S / dt$) and is in general as follows,

$$\mu = \frac{d}{dt}((t/\alpha)^{\beta(t)}) \text{ or } \mu = (t/\alpha)^{\beta(t)} \times [\frac{\beta(t)}{t} + \ln(t/\alpha) \times \frac{d\beta(t)}{dt}] \qquad (2)$$

where $\mu$ is the mortality rate (denoted as hazard rate or force of mortality), meaning the relative rate for the survival function decrease. Obviously, mortality trends should be directly associated with shape parameter trends. It is noteworthy that 'age dependence of the shape parameter' intrinsically makes the mortality function inevitably complex and difficult for modeling.

Conveniently, the value of characteristic life ($\alpha$) is always found at the duration for survival to be 'exp(-1)'; this is known as the characteristic life. This feature gives the advantage of looking for the value of $\alpha$ simply by a graphical analysis of the survival curve. In turn, with the observed value of $\alpha$, we can plot 'rectangularity' with age by the mathematical equivalence of '$\beta(t)=\ln(-\ln S)/\ln(t/\alpha)$'. If $\beta(t)$ is not constant with age, this obviously implies that '$\beta(t)$ is a function of age'. On the other hand, $\beta(t)$ mathematically approaches infinity as the age $t$ approaches the value of $\alpha$ or the denominator '$\ln(t/\alpha)$' approaches zero. This feature of $\beta(t)$ can leave 'traces of $\alpha$' in the plot of $\beta(t)$, so we can observe variations of $\beta(t)$ and $\alpha$ at once in the plot of the shape parameters. If $\beta(t)$ (except for the mathematical singularity (traces of $\alpha$)) can be expressed by an adequate mathematical function, the survival and mortality functions can be calculated by the mathematically expressed $\beta(t)$. Only two parameters, $\beta(t)$ and $\alpha$, determine the survival and mortality functions. In empirical practice, we could use a linear expression for the mature phase (middle age) and a quadratic expression for senescence phase (from characteristic life to maximum age): $\beta(t) = \beta_0 + \beta_1 t + \beta_2 t^2 + ...$, where the associated coefficients were determined by a regression analysis in the plot of shape parameter curve. And thus, the derivative of $\beta(t)$ was obtained as follows: $d\beta(t)/dt = \beta_1 + 2\beta_2 t + ...$, which is the important finding in the previous papers (Weon 2004a, 2004b), that is, the shape parameter for humans is a function of age.

## 3. An approximate relationship between the mortality rate and the shape parameter

We are able to observe a certain degree of universality for the linear progression of the shape parameter for ages earlier than characteristic life in many modern developed countries. Figure 1A shows that in thirteen developed countries (Austria 1999, Canada 1996, Denmark 2000, England & Wales 1998, Finland 2000, Italy 1999, Germany 1999, Japan 1999, Netherlands 1999, Norway 2000, Sweden 2001, Switzerland 2001 and USA 1999), the shape parameter for ages 0 up to 80 was estimated to be $\beta(t)=1.44578+0.08581t$. It was shown in the previous paper (Weon 2004a), that these countries have a characteristic life of more than 80 years. Using this universality of $\beta(t)$, we could simulate the relationship of the mortality deceleration and the shape parameter in the following way.

For our analysis, we separately evaluate the impact of the mathematical terms on the mortality rate as follows,

$$\mu = A \times B; \text{ where } A = (t/\alpha)^{\beta(t)} \text{ and } B = [\frac{\beta(t)}{t} + \ln(t/\alpha) \times \frac{d\beta(t)}{dt}] \qquad (3)$$

In Fig. 1B, since the term *B* is almost invariant after adulthood (> ~30), the result is that we could obtain the approximate relationship of the mortality rate (log-scale) and the shape parameter (linear-scale) after adulthood.

$$\ln \mu \propto \beta(t) \tag{4}$$

This relationship implies that the mortality rate would follow the Gompertz law (exponential growth in the mortality rate with age), when the shape parameter is linearly proportional to age, but the mortality rate would deviate from the Gompertz law when the shape parameter is bent (non-linear) with age.

Furthermore, the age dependence of the shape parameter, $\beta(t)$, for population dynamics may be a general law that includes the Gompertz model and the Weibull model: i) the Gompertz model when $\beta(t)$ is linearly proportional to age and ii) the Weibull model when $\beta(t)$ is a constant. The Gompertz model (Gompertz 1825) and the Weibull model (Weibull 1951) are the most generally used models at present (Gavrilov and Gavrilova 2001). Interestingly, the Gompertz model is more commonly used to describe biological systems, whereas the Weibull model is more commonly applicable to technical devices (Gavrilov and Gavrilova 2001). Particularly for aging patterns, it is the shape parameter that distinguishes humans from technical devices. It seems to show the difference between humans and technical devices in terms of '*robustness*'. The fundamental difference for robustness between biological systems and technical devices is obvious (Gavrilov and Gavrilova 2001). In the previous papers (Weon 2004a, 2004b), the age-dependent shape parameter is changed from approximately 0.5 to 10 with age for the typical human survival curves. This feature is in great contrast to technical devices typically having a constant shape parameter (Nelson 1990). We attribute the age dependence of the shape parameter to the resistance to aging, which can be described as a struggle to extend the duration of life by increasing the shape parameter, or the *homeostasis* and the *adaptation* of biological systems, which must have the homeostasis and the adaptation to maintain stability and to survive (Weon 2004a).

For humans over much of the age range, the Gompertz model still gives an excellent approximation. At the high ages, however, the law does not apply very well (Thatcher, Kannisto, and Vaupel 1998). In the Weon model, a linear expression for the shape parameter is appropriate before characteristic life and a quadratic expression is appropriate after characteristic life for the modern demographic curves (Weon 2004a). The age-dependent shape parameter in the Weon model can be seen as a measure of the deviation from the Gompertz law at senescence and the mortality deceleration at high ages.

## 4. A mathematical limit of longevity by mortality decrease

In order to model the mortality rate after characteristic life (or at high ages), we should model the shape parameter accurately. For example in Fig. 2A, we plotted $\beta(t)$ after the characteristic life in Switzerland (1876-2001); sometimes we had to omit several points close to the characteristic life to ensure the quality of the regression analysis. Also, we accomplished the regression analysis to confirm the best mathematical model for $\beta(t)$. In this case, we used a quadratic expression for $\beta(t)$. In this way, in Fig. 2B, we examined the demographic data from Sweden (1861-2001). The detailed results of the regression analysis are recorded in Table 1 (Switzerland) and Table 2 (Sweden).

The mathematical model for $\beta(t)$ enables us to model the mortality rate after characteristic life. We are able to see the example of modeling $\mu$ through modeling $\beta(t)$ for 2001 and 1980 for Switzerland in Fig. 3A and 3B. It is especially noted that in Fig. 3B, the mortality rate shows the decrease after a plateau around ages 110-115 and this shows the emergence of the mathematical limit around ages 120-130.

The following way is convenient to evaluate the mathematical limit. The mortality rate should be mathematically positive ($\mu > 0$); since the term $A$ is always positive ($A > 0$), thus the term $B$ should be positive ($B > 0$). Therefore, the criterion for the mathematical limit of longevity, implying the ultimate limit of longevity which is able to be determined by the mortality dynamics, can be given by,

$$\frac{d\beta(t)}{dt} > -\frac{\beta(t)}{t \ln(t/\alpha)} \tag{5}$$

For analysis, the left term denotes the term $C$ and the right term denotes the term $D$. In order to evaluate the mathematical limit of longevity, the age-dependent shape parameter after characteristic life must be accurately modelled. In the previous paper (Weon 2004a), we successfully used a quadratic expression ($\beta(t)=\beta_0+\beta_1 t+\beta_2 t^2$) for the estimation of the shape parameter after characteristic life in the case of the years of 1876-2001 in Switzerland.

Interestingly, the quadratic coefficient ($\beta_2$) supports the bending of the shape parameter, which would be directly associated with the mortality deceleration as discussed above. Furthermore, the quadratic coefficient ($\beta_2$) is important to evaluate the mathematical limit of longevity, since it determines the slope with age in the derivative ($\beta_1+2\beta_2 t$) of the quadratic expression of the shape parameter or in the term $C$.

## 5. Demographic evidence in Switzerland and Sweden

Demographic data, the period life tables (all sexes, 1x1) for the years of 1876-2001 in Switzerland and for the years of 1861-2001 in Sweden, were taken from the Human Mortality Database (http://www.mortality.org). We calculated the trend lines for the shape parameters after characteristic life in Table 1 (Switzerland) and Table 2 (Sweden) by the quadratic expression for the shape parameter. We could see that the estimated quadratic expression for the shape parameter is very accurate according to the high value of the coefficient of determinant, $R^2$, and the extremely low $P$-value (<0.0001).

In order to evaluate the mathematical limits of recent decades in Switzerland and Sweden, we separately calculate the term $C$ and the term $D$, and then plot the mathematical terms as a function of age in Fig. 4A (Switzerland) and 4B (Sweden). The term $C$ decreases with increasing age, whereas the term $D$ increases with increasing age. Interestingly, the term $C$ tends to become steeper over time in recent decades. The slope of the term $C$ is important to evaluate the mathematical limit of longevity. The most significant observation is that the mathematical limit of longevity becomes shorter as the slope of the term $C$ becomes steeper in Fig. 4A and 4B. This phenomenon, as shown in Fig. 5, indicates that the bending of the shape parameter (or the quadratic coefficient ($\beta_2$)) is associated with the mathematical limit of longevity, which is due to the mortality decrease at the highest ages. The slope of the term $C$ or the quadratic coefficient ($β_2$) tends to increase in recent decades as shown in Fig. 4A and 4B, and in Table 1 and Table 2.

**6. Demographic evidence in the other eleven countries**

The findings from Switzerland and Sweden are very interesting. We wish to expand our investigation to the other eleven developed countries (Austria 1999, Canada 1996, Denmark 2000, England & Wales 1998, Finland 2000, Italy 1999, Germany 1999, Japan 1999, Netherlands 1999, Norway 2000, Sweden 2001, Switzerland 2001 and USA 1999) as the most recent trends. Demographic data, the period life tables (all sexes, 1x1) for the most recent years between 1996-2001 in the eleven developed countries, were taken from the Human Mortality Database (http://www.mortality.org). We calculated the trend lines for the shape parameters after characteristic life in Table 3. We could see that the estimated quadratic expression for the shape parameter is very accurate according to the high value of the coefficient of the determinant, $R^2$, and the extremely low *P*-value (<0.0001).

We accomplished the similar examinations in Fig. 6, 7, and 8. We found out that the quadratic expression for $β(t)$ is appropriate after characteristic life in Fig. 6 for the eleven countries. We evaluated the mathematical limits of longevity in Fig. 7. Finally, we could obtain the similar phenomenon between the mathematical limits and the quadratic coefficient (the bending of the shape parameter) in Fig. 8. These results confirm that the bending of the shape parameter after characteristic life is correlated with the mortality deceleration (or decrease), and consequently, is associated with the mathematical limit of longevity. The findings suggest that the mathematical limit of longevity can de induced by the mortality deceleration (or decrease) in nature.

**7. Discussion**

Vaupel et al. has suggested that the human mortality rates could decrease after having reached a maximum (a plateau or a ceiling of mortality) (Vaupel et al. 1998). The results reported by Robine and Vaupel strongly support the finding that the mortality rate does not increase according to the Gompertz curve at the highest ages and the results are consistent with a plateau around age 110-115, and their earlier study suggests that mortality may fall after age 115 (Robine and Vaupel 2002). In humans, the deceleration in the mortality rate is not seen until after 80 years of age, but a clear deviation from the predicted Gompertz model of exponential increases in mortality is observed. If individuals such as Jeanne Calment, who lived to the ripe old age of 122 years and 164 days, are included, then mortality rates are seen to decrease with age after 110 years of age (Helfand and Inouye 2002). The Weon model is well able to describe the mortality decrease after the mortality plateau, including the mortality deceleration, and further, the model predicts the mathematical limit of longevity by the mortality decrease. It may be arguable to evaluate the mathematical limit of longevity by the mortality decrease based on this assumption, which fit the available data from characteristic life (around age 80) up to approaching age 110, and the Weon model will continue to apply up to older age. However, the Weon model for mortality rates from the estimation of the shape parameter is very reliable in this investigation, so we believe that such evaluations are reasonable.

Yet to our surprise, the trend of the mathematical limit of longevity in recent decades seems to be contrary to all our knowledge. We should note that the characteristic life in both of Switzerland and Sweden has increased constantly for more than a century in Fig. 9, which is consistent with other literature (Wilmoth et al. 2000,

Oeppen and Vaupel 2002, Robine and Vaupel 2002). On the contrary, the mathematical limits of longevity by the mortality decrease seem to tend to decrease in recent decades. Furthermore, the mathematical limit of longevity in the thirteen developed countries approaches approximately 120 to 130 years in recent decades in Fig. 4 and 5, and Fig. 7 and 8. Interestingly, the bending of the shape parameter tends to be more intensive with increasing characteristic life in the previous paper (see Fig. 6 in Weon 2004a). This suggests that there probably is an inherent correlation between the mortality deceleration (or decrease) and the increase of characteristic life (or life expectancy). The mathematical limit of longevity may be associated with the fundamental mechanisms of human longevity. We intend to investigate this question in subsequent research.

## 8. Summary

The age-dependent shape parameter can be a measure of the deviation from the Gompertz law at senescence or the mortality deceleration (or decrease) at the highest ages. According to the Weon model, we are well able to describe the mortality decrease after the mortality plateau, including the mortality deceleration. Furthermore, we are able to simply define the mathematical limit of longevity by the mortality decrease. From the demographic analysis of the historical trends in Switzerland (1876-2001) and Sweden (1861-2001), and the most recent trends in the other eleven developed countries, we confirm that the bending of the shape parameter after characteristic life is correlated with the mortality deceleration (or decrease) and consequently, it is associated with the mathematical limit of longevity. The results suggest that the mathematical limit of longevity can be induced by the mortality deceleration (or decrease) in nature. These findings will give us a breakthrough to study the mortality dynamics at the highest ages.

**Acknowledgements**

I am grateful to the Human Mortality Database (Dr. John R. Wilmoth, as a director, in The University of California, Berkeley and Dr. Vladimir Shkolnikov, as a co-director, in The Max Planck Institute for Demographic Research) for allowing me to access the demographic data for this research.

**Figures**

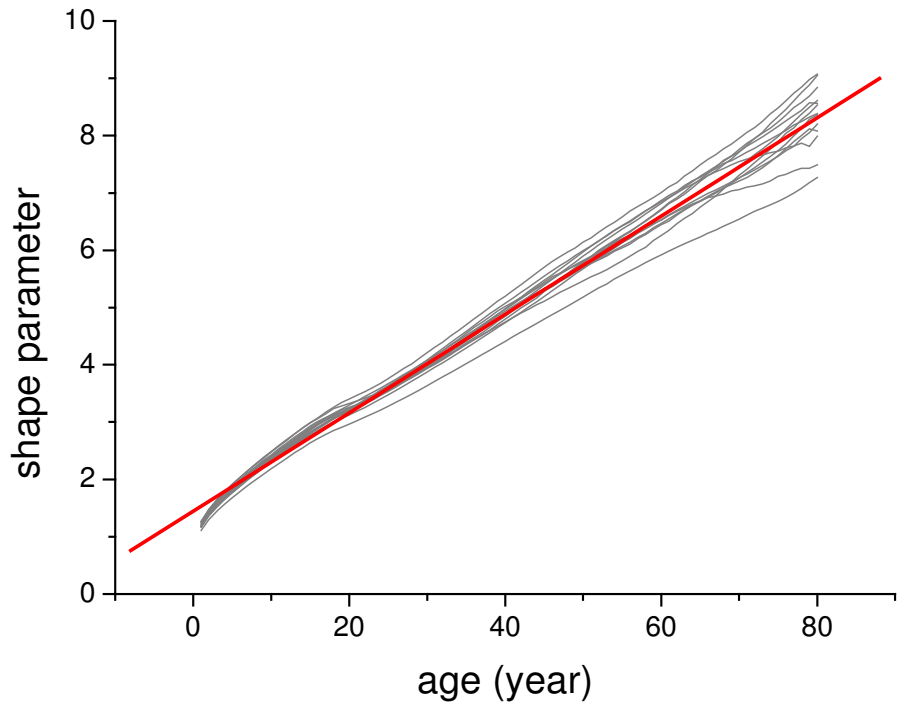

Fig. 1A. Universality of linear progression of the shape parameter for ages 0-80 in thirteen developed countries (1996-2001).

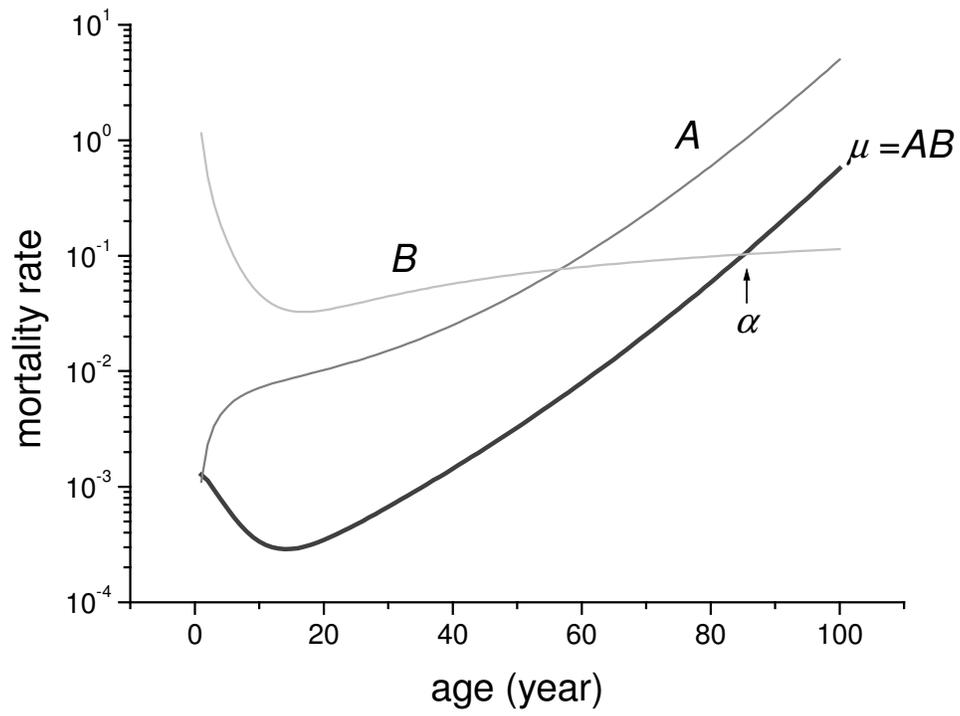

Fig. 1B. Simulation of the mortality rate through the universality of the shape parameter.

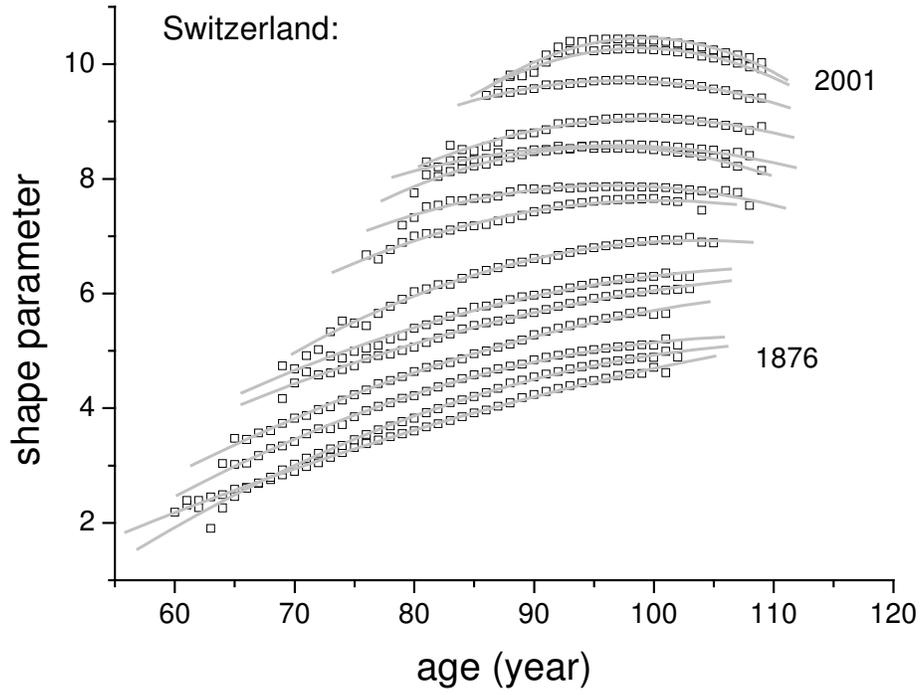

Fig. 2A. Historical trends of the shape parameters after characteristic life in Switzerland (1876-2001).

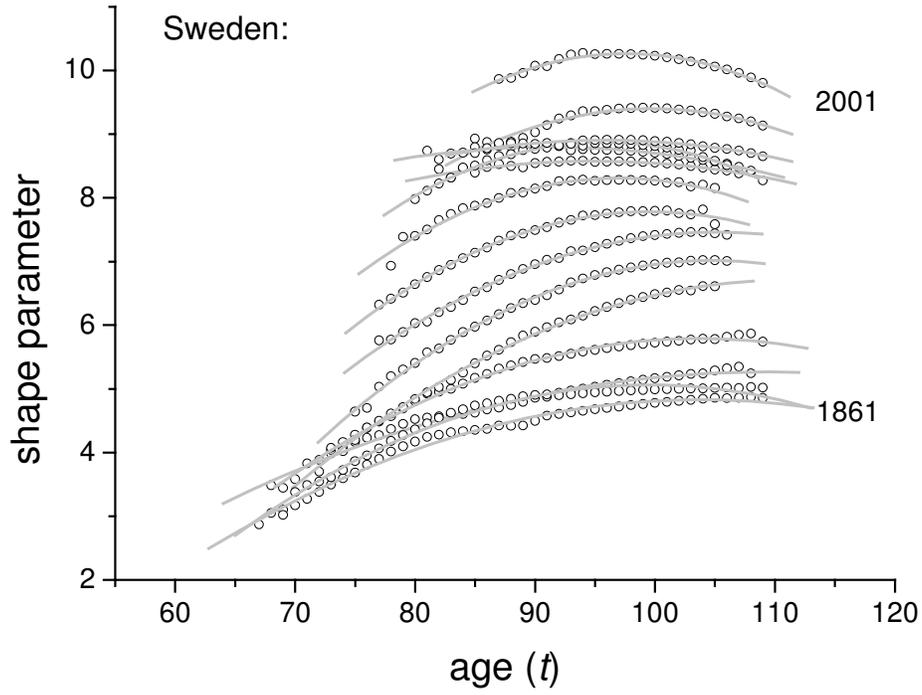

Fig. 2B. Historical trends of the shape parameters after characteristic life in Sweden (1861-2001).

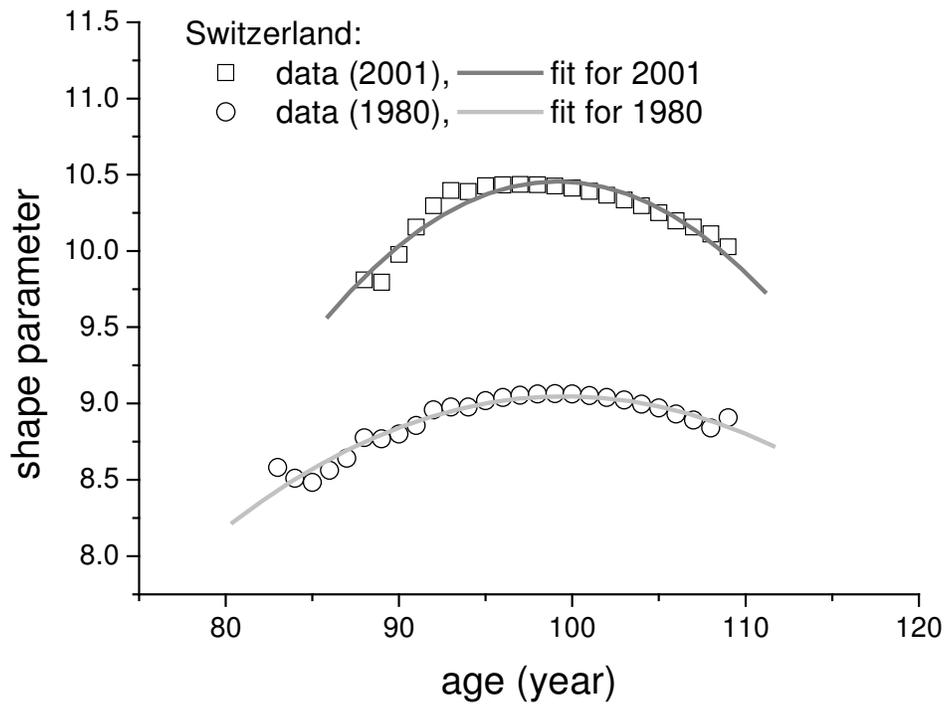

Fig. 3A. Modeling the shape parameter as a quadratic expression for 1980 and 2001 for Switzerland.

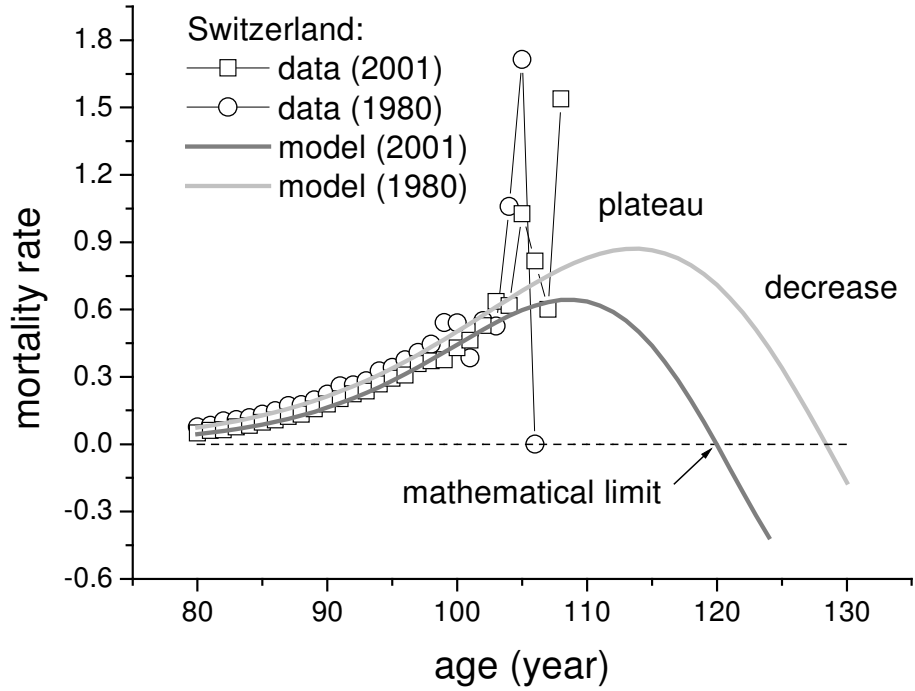

Fig. 3B. Modeling the mortality rate through modeling the shape parameter and the death rate (mortality) data for 2001 and 1980 for Switzerland.

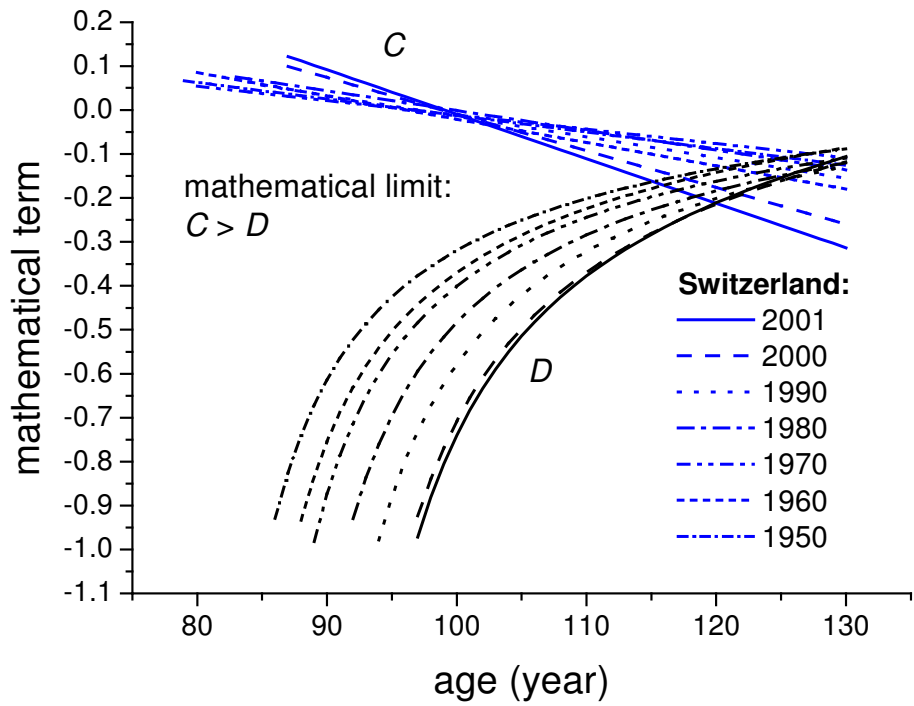

Fig. 4A. Calculations of mathematical terms, *C* and *D*, as a function of age for Switzerland.

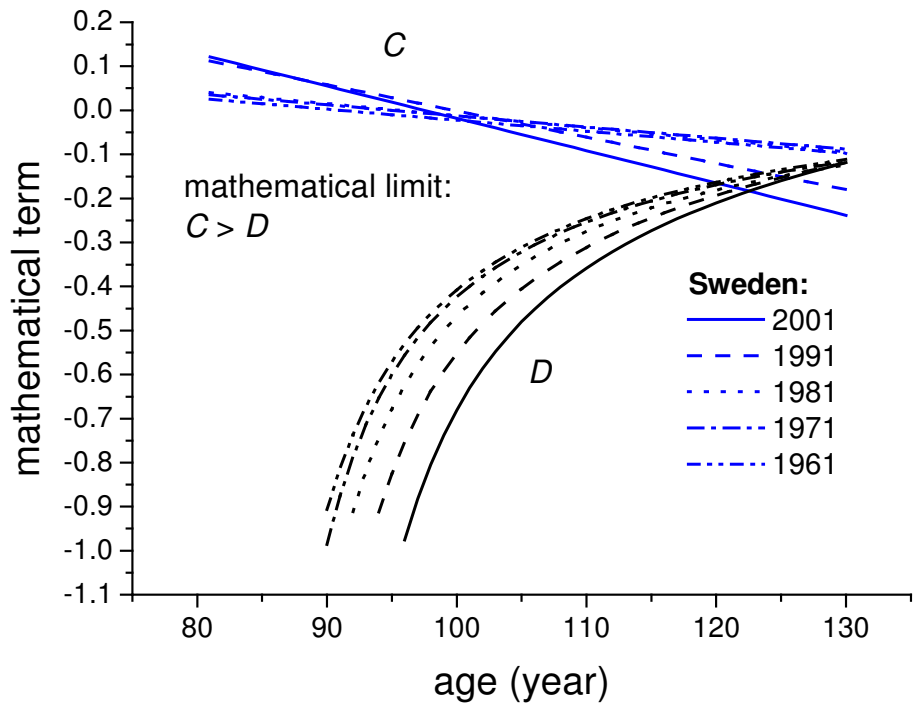

Fig. 4B. Calculations of mathematical terms, *C* and *D*, as a function of age for Sweden.

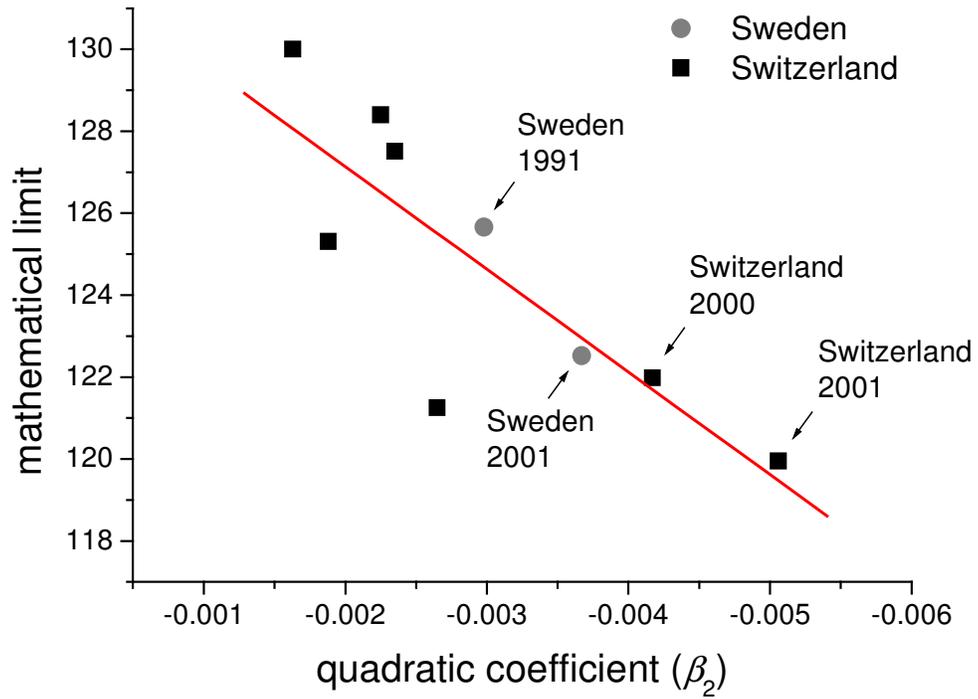

Fig. 5. Shortening mathematical limits of longevity with increasing the quadratic coefficient ($\beta_2$) in both of Switzerland and Sweden.

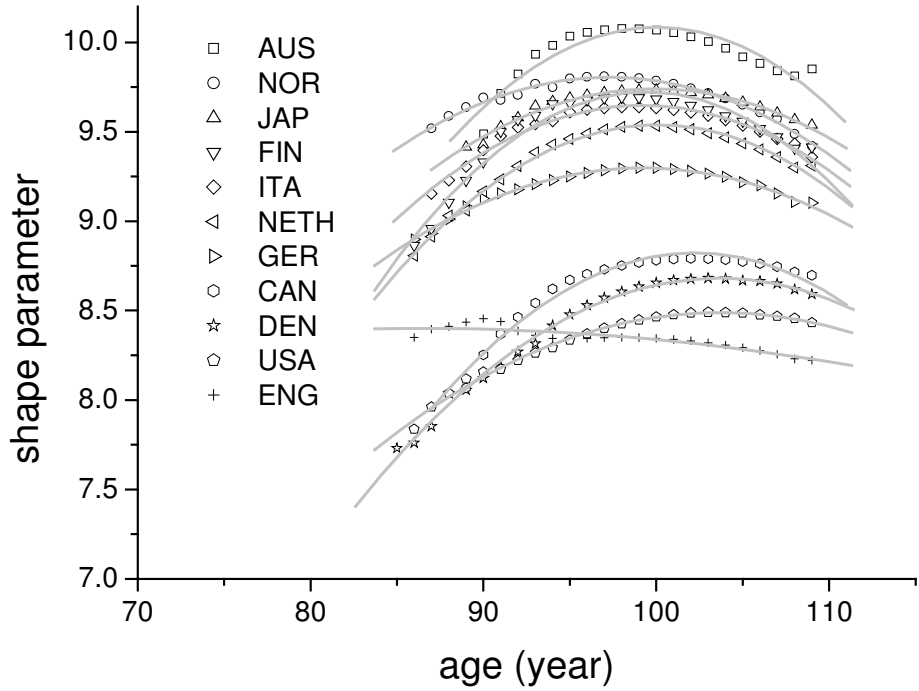

Fig. 6. Trends of the shape parameters after characteristic life in the other eleven developed countries (1996-2001) as the most recent trends. Note: AUS-Austria, NOR-Norway, JAP-Japan, FIN-Finland, ITA-Italy, NETH-Netherlands, GER-Germany, CAN-Canada, DEN-Denmark, USA-Unites State of America, ENG-England & Wales.

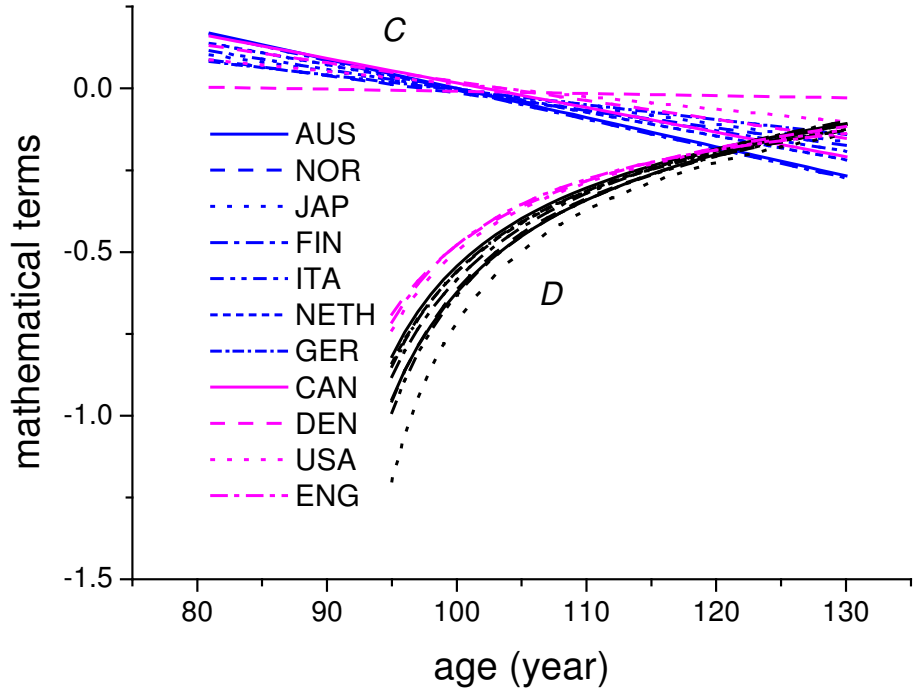

Fig. 7. Calculations of mathematical terms, *C* and *D*, as a function of age in the eleven developed countries. Note: AUS-Austria, NOR-Norway, JAP-Japan, FIN-Finland, ITA-Italy, NETH-Netherlands, GER-Germany, CAN-Canada, DEN-Denmark, USA-Unites State of America, ENG-England & Wales.

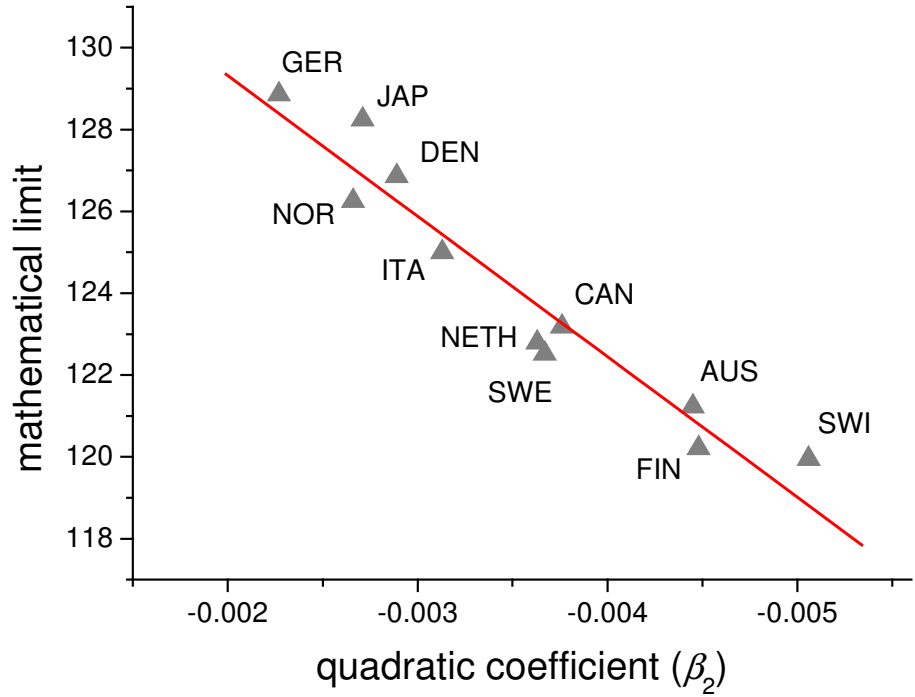

Fig. 8. Shortening mathematical limits of longevity with increasing the quadratic coefficient ($β_2$) in the eleven developed countries including Switzerland and Sweden. Note: AUS-Austria, NOR-Norway, JAP-Japan, FIN-Finland, ITA-Italy, NETH-Netherlands, GER-Germany, CAN-Canada, DEN-Denmark, USA-Unites State of America, ENG-England & Wales.

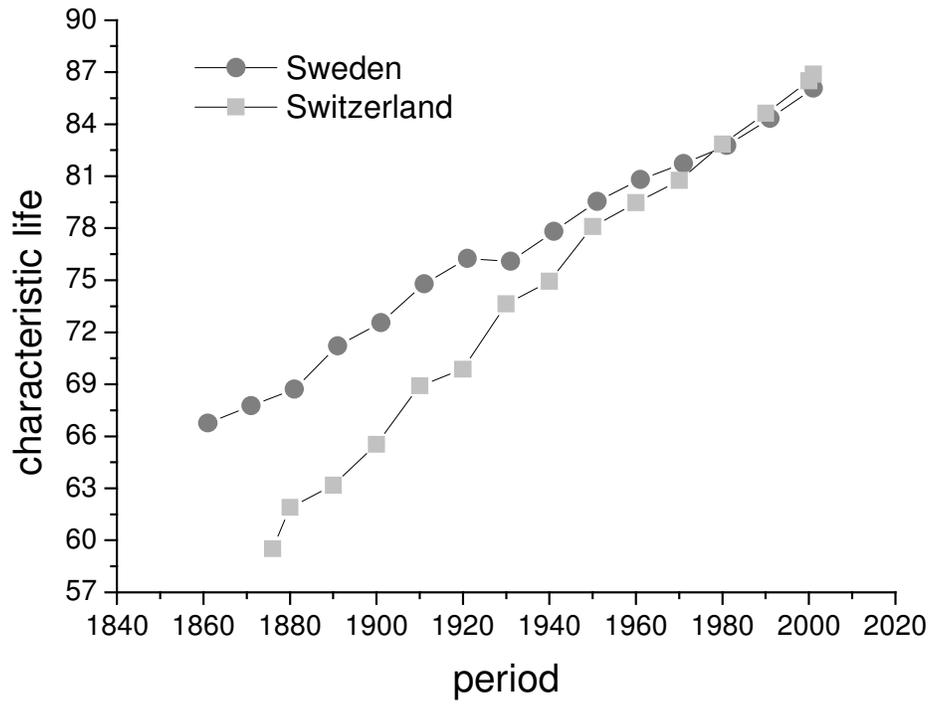

Fig. 9. Historical trends of characteristic life in both of Switzerland and Sweden.

Table 1. Shape parameter estimated as a quadratic expression after characteristic life for 1876-2001 for Switzerland.

| Year | $\beta_0$ | $\beta_1$ | $\beta_2$ | $R^2$ | (P-value) |
|------|-----------|-----------|-----------|---------|-----------|
| 2001 | -39.22154 | 1.00226   | -0.00506  | 0.90494 | (<0.0001) |
| 2000 | -30.49174 | 0.82469   | -0.00417  | 0.93217 |           |
| 1990 | -12.42385 | 0.45586   | -0.00235  | 0.97375 |           |
| 1980 | -13.29566 | 0.44890   | -0.00225  | 0.93770 |           |
| 1970 | -6.617620 | 0.31492   | -0.00163  | 0.84224 |           |
| 1960 | -15.90502 | 0.50906   | -0.00265  | 0.95876 |           |
| 1950 | -9.649830 | 0.36351   | -0.00188  | 0.92426 |           |
| 1940 | -9.475550 | 0.34073   | -0.00170  | 0.97375 |           |
| 1930 | -11.57780 | 0.35689   | -0.00172  | 0.99198 |           |
| 1920 | -6.145470 | 0.22396   | -0.00099  | 0.97999 |           |
| 1910 | -4.798780 | 0.18597   | -0.00077  | 0.98526 |           |
| 1900 | -6.486440 | 0.20651   | -0.00085  | 0.99744 |           |
| 1890 | -8.211080 | 0.24414   | -0.00111  | 0.99660 |           |
| 1880 | -9.158820 | 0.25042   | -0.00110  | 0.99056 |           |
| 1876 | -4.318390 | 0.13551   | -0.00045  | 0.99815 |           |

Note: $\beta(t) = \beta_0 + \beta_1 t + \beta_2 t^2$

Table 2. Shape parameter estimated as a quadratic expression after characteristic life for 1861-2001 for Sweden.

| Year | $\beta_0$ | $\beta_1$ | $\beta_2$ | $R^2$ | (P-value) |
|------|-----------|-----------|-----------|---------|-----------|
| 2001 | -24.68505 | 0.71594 | -0.00367 | 0.96717 | (<0.0001) |
| 1991 | -20.28298 | 0.59479 | -0.00298 | 0.91251 | |
| 1981 | -3.653530 | 0.26206 | -0.00137 | 0.88759 | |
| 1971 | -2.792430 | 0.23941 | -0.00126 | 0.84286 | |
| 1961 | -1.574590 | 0.22763 | -0.00125 | 0.91608 | |
| 1951 | -22.55137 | 0.65682 | -0.00343 | 0.98771 | |
| 1941 | -21.82020 | 0.62146 | -0.00320 | 0.97352 | |
| 1931 | -21.50172 | 0.58798 | -0.00295 | 0.99537 | |
| 1921 | -17.87890 | 0.48209 | -0.00229 | 0.99670 | |
| 1911 | -22.02449 | 0.55444 | -0.00265 | 0.99770 | |
| 1901 | -17.00765 | 0.42768 | -0.00193 | 0.99772 | |
| 1891 | -14.30066 | 0.38774 | -0.00187 | 0.99151 | |
| 1881 | -14.62550 | 0.39595 | -0.00199 | 0.98411 | |
| 1871 | -6.756830 | 0.22012 | -0.00101 | 0.98537 | |
| 1861 | -10.20184 | 0.28998 | -0.00140 | 0.98905 | |

Note: $\beta(t) = \beta_0 + \beta_1 t + \beta_2 t^2$

Table 3. Shape parameter estimated as a quadratic expression after characteristic life for the eleven developed countries.

| Country | $\beta_0$ | $\beta_1$ | $\beta_2$ | $R^2$ | (P-value) |
|---|---|---|---|---|---|
| Austria | -34.43938 | 0.89024 | -0.00445 | 0.84619 | (<0.0001) |
| Norway | -15.38753 | 0.51811 | -0.00266 | 0.98490 | |
| Japan | -17.30636 | 0.54138 | -0.00271 | 0.97326 | |
| Finland | -34.59301 | 0.89137 | -0.00448 | 0.95626 | |
| Italy | -21.20022 | 0.62189 | -0.00313 | 0.98166 | |
| Netherlands | -26.79699 | 0.72611 | -0.00363 | 0.99124 | |
| Germany | -13.02618 | 0.44991 | -0.00227 | 0.99067 | |
| Canada | -30.47682 | 0.76892 | -0.00376 | 0.96768 | |
| Denmark | -22.35540 | 0.59909 | -0.00289 | 0.99459 | |
| USA | -12.23463 | 0.39989 | -0.00193 | 0.99228 | |
| England & Wales | +5.92728 | 0.05735 | -0.00033 | 0.82539 | |

Note: $\beta(t) = \beta_0 + \beta_1 t + \beta_2 t^2$